\documentclass[preprint,12pt, a4paper]{elsarticle}

\usepackage{amssymb}
\usepackage{amsmath}
\usepackage{hyperref}
\usepackage{float}
\usepackage{adjustbox}

\usepackage{booktabs} 
\usepackage{xcolor}
\usepackage{caption} 
\usepackage{hyperref}
\hypersetup{
    colorlinks=true,
    linkcolor=blue,
    filecolor=blue,
    urlcolor=blue,
    citecolor=blue
}

\usepackage{listings}

\usepackage[T2A,T1]{fontenc}
\usepackage[utf8]{inputenc}
\usepackage[ukrainian,english]{babel}

\usepackage{geometry}
\usepackage{tabularx}
\usepackage{ragged2e}

\journal{SoftwareX}

\begin{document}

\renewcommand{\labelenumii}{\arabic{enumi}.\arabic{enumii}}
\definecolor{codegreen}{RGB}{0,128,0}
\definecolor{codepink}{RGB}{219,0,219}
\definecolor{codegray}{RGB}{248,248,248}

\lstset{
    backgroundcolor=\color{codegray},
    basicstyle=\ttfamily\small,
    numbers=left,
    numberstyle=\tiny\color{gray},
    keywordstyle=\color{codegreen},
    stringstyle=\color{codepink},
    commentstyle=\color{codegreen},
    breaklines=true,
    showstringspaces=false,
    frame=single,
    rulecolor=\color{black},
    language=Python,
    captionpos=b
}

\begin{frontmatter}

\title{BlockingPy: approximate nearest neighbours for blocking of records for entity resolution}

\author[label1,label2]{Tymoteusz Strojny
% \fnref{label4}
}
\author[label3,label2]{Maciej Beręsewicz}
\address[label1]{Faculty of Mathematics and Computer Science, Adam Mickiewicz University in Poznań, Uniwersytetu Poznańskiego 4, 61-614 Poznań, Poland}
\address[label2]{Centre for the Methodology of Population Studies, Statistical Office in Poznań, Wojska Polskiego 27/29, 60-624 Poznań, Poland}
\address[label3]{Department of Statistics, Poznań University of Economics and Business, Al. Niepodległości 10, 61-875 Poznań,  Poland, maciej.beresewicz@ue.poznan.pl}

\begin{abstract}
Entity resolution (probabilistic record linkage, deduplication) is a key step in scientific analysis and data science pipelines involving multiple data sources. The objective of entity resolution is to link records without common unique identifiers that refer to the same entity (e.g., person, company). However, without identifiers, researchers need to specify which records to compare in order to calculate matching probabilities and reduce computational complexity. One solution is to deterministically block records based on some common variables, such as names, dates of birth, or sex, or by using phonetic algorithms. However, this approach assumes that these variables are free of errors and completely observed, which is often not the case. To address this challenge, we have developed a Python package, \texttt{BlockingPy}, which relies on schema-agnostic blocking using modern approximate nearest neighbour search and graph algorithms to reduce the number of comparisons. The package supports both CPU and GPU execution. In this paper, we present the design of the package, its functionalities and two case studies related to official statistics. The presented software will be useful for researchers interested in linking data from various sources.
\end{abstract}


\end{frontmatter}

%\linenumbers

\clearpage
\section*{Metadata}\label{sec-metadata}

\begin{table}[!h]
\begin{tabular}{|l|p{6.5cm}|p{6.5cm}|}
\hline
\textbf{Nr.} & \textbf{Code metadata description} & \textbf{Metadata} \\
\hline
C1 & Current code version & v0.2.8 \\
\hline
C2 & Permanent link to code/repository used for this code version & N/A \\
\hline
% \url{https://github.com/ncn-foreigners/BlockingPy}
C3  & Permanent link to Reproducible Capsule & \url{https://blockingpy.readthedocs.io/en/latest/examples/record_linkage.html}\\
\hline
C4 & Legal Code License   & MIT License \\
\hline
C5 & Code versioning system used & Git \\
\hline
C6 & Software code languages, tools, and services used & Python \\
\hline
C7 & Compilation requirements, operating environments \& dependencies & 
pandas, numpy, scipy, annoy, hnswlib, pynndescent, scikit-learn, igraph, nltk, voyager, faiss-cpu, faiss-gpu-cuvs, mlpack, model2vec \\
\hline
C8 & If available Link to developer documentation/manual & \url{https://blockingpy.readthedocs.io/en/latest/index.html} \\
\hline
C9 & Support email for questions & tymoteusz.strojny@gmail.com\\
\hline
\end{tabular}
\caption{Code metadata (mandatory)}
\label{codeMetadata} 
\end{table}

% \textit{Optionally, you can provide information about the current executable
% software version filling in the left column of
% Table~\ref{executabelMetadata}. Please leave the first column as it is.}

% \begin{table}[!h]
% \begin{tabular}{|l|p{6.5cm}|p{6.5cm}|}
% \hline
% \textbf{Nr.} & \textbf{(Executable) software metadata description} & \textbf{Please fill in this column} \\
% \hline
% S1 & Current software version & For example 1.1, 2.4 etc. \\
% \hline
% S2 & Permanent link to executables of this version  & For example: \url{https://github.com/combogenomics/DuctApe/releases/tag/DuctApe-0.16.4} \\
% \hline
% S3  & Permanent link to Reproducible Capsule & \\
% \hline
% S4 & Legal Software License & List one of the approved licenses \\
% \hline
% S5 & Computing platforms/Operating Systems & For example Android, BSD, iOS, Linux, OS X, Microsoft Windows, Unix-like , IBM z/OS, distributed/web based etc. \\
% \hline
% S6 & Installation requirements \& dependencies & \\
% \hline
% S7 & If available, link to user manual - if formally published include a reference to the publication in the reference list & For example: \url{http://mozart.github.io/documentation/} \\
% \hline
% S8 & Support email for questions & \\
% \hline
% \end{tabular}
% \caption{Software metadata (optional)}
% \label{executabelMetadata} 
% \end{table}

\section{Motivation and significance}

Entity resolution requires statistical and computational methodologies to accurately identify matching records across datasets without unique identifiers. This process underpins countless research endeavours in disciplines such as epidemiology, economics, official statistics and historical research where data integration is fundamental to scientific discovery \cite{binette2022almost, tancredi2020unified}. Traditional blocking techniques, while computationally efficient, exhibit limitations when confronted with various data quality issues such as missing values, transliteration, typographical errors, and inconsistent formatting. Existing deterministic blocking methods rely on user-defined exact-matching rules that assume key fields (e.g., name, date of birth) are error-free, potentially leading to missed links and false matches \cite{Christen2012, Herzog2007, jugl2024gecko}. An alternative approach is schema-agnostic blocking \cite{Papadakis2015, Thirumuruganathan2021,Papadakis2020}, which concatenates all attribute values of a~particular entity into a~single textual representation before applying blocking (and can be combined with pre-trained language models).

The scientific community has developed several tools to address record linkage, including R packages such as \texttt{RecordLinkage} \cite{RecordLinkage2010}, \texttt{reclin2} \cite{van2022reclin2}, \texttt{fastLink} \cite{Enamorado2019fastLink}, or Python implementations like \texttt{splink} \cite{Splink}, \texttt{recordlinkage} \cite{recordlinkage}, \texttt{Dedupe} \cite{dedupe}, \texttt{pyJedAI} \cite{pyJedAI22}, or \texttt{FEBRL} \cite{christen2008febrl}. However, these solutions typically rely on user-defined deterministic rules that may miss certain groups of records or result in large blocks of records when data irregularities are present. 

Some solutions to these issues were proposed in the literature, for instance, the use of the SOUNDEX algorithm and other phonetic algorithms to block records on character strings that sound similar phonetically \cite{christen2012data}; blocking using the Locality Sensitive Hashing (LSH; \cite{klsh}), an approximate nearest neighbour (ANN) algorithm to search for similar records (see, e.g., the \texttt{klsh} package in R) or the use of Bloom filters, useful in particular for privacy-preserving record linkage (see, \cite{christen2020linking}; for implementation see \texttt{blocklib} in Python).

The \texttt{BlockingPy} package relies on the schema-agnostic approach and uses state-of-the-art ANN search algorithms and graph-based indexing techniques, which allow it to work even with imperfect data. \texttt{BlockingPy} implements an ANN-based blocking procedure ($n$-gram encoding or embeddings, neighbour search, graph components) behind a single \texttt{pandas}-style call, cutting custom rule-writing and setup time.

The structure of the paper is as follows. In Section \ref{sec-blocking}, we briefly review the blocking procedures in existing Python packages. In Section \ref{sec-software}, we provide details regarding the implementation and the API. In Section \ref{sec-examples}, we provide two examples of probabilistic record linkage and deduplication. The paper concludes with a brief review of the possible impact and conclusions.

\section{Existing software covering blocking procedures}\label{sec-blocking}

This section focuses on Python packages (\texttt{splink}, \texttt{recordlinkage}, and \texttt{blocklib}), but readers interested in deterministic blocking procedures are referred to the \texttt{fastLink} and \texttt{reclin2} packages in R. For a~comparison of Python packages, see Table~\ref{tab:blocking_methods} in the Appendix.

The \texttt{splink} package allows for deterministic blocking rules using the function \texttt{block\_on()} which specifies either the variables (e.g., \texttt{block\_on("city")}) or more advanced combinations (e.g., \texttt{block\_on("substr(first\_name, 1,1)", "surname")}), which then are translated to a SQL query for the DuckDB (via the \texttt{BlockingRuleCreator} class). More advanced users may be interested in creating their own rules, which is possible through the \texttt{CustomRule} class. \texttt{Splink} allows users to specify the list of blocking rules for the \texttt{Linker} class. \texttt{Splink} allows users to inspect the blocking via the \texttt{splink.blocking\_analysis} method.

Additionally, \texttt{splink} supports blocking using phonetic algorithms such as Soundex, Metaphone, or Double Metaphone either by applying them directly within the blocking rules (e.g., \texttt{block\_on("soundex(surname)")}), or by preprocessing columns to store phonetic encodings separately. In online documentation we show that using \texttt{splink} with  \texttt{BlockingPy} together results in significant improvements in the performance metrics by capturing comparison pairs that would otherwise be missed.

\texttt{recordlinkage} package supports deterministic blocking via the \texttt{record\-linkage\-.Index} class and its methods. This package allows for deterministic blocking using \texttt{BlockIndex()} and \texttt{SortedNeighbourhood()} where specification of blocking variables should be provided by a~vector of column names and optionally \texttt{window} for the \texttt{SortedNeighbourhood()}, which accounts for multiple mistakes. 

% The only software that provides similar functionality is the \texttt{blocklib} package (cf. \cite{zhang2018scalable, Anonlink}) and \texttt{vicinity} \cite{vicinity}. The first uses probability signature and LSH-based Lambda-Fold Redundant as described in \cite{zhang2018scalable}. The second uses \texttt{semhash} \cite{minishlab2025semhash}, which allows deduplication of records using kNN-based embedding comparisons and offers exact-match blocking. In contrast, \texttt{BlockingPy} allows various inputs as well as ANN algorithms and graphs, focusing on providing accurate blocking for probabilistic record linkage and deduplication. In Appendix \ref{appen-blocklib}, we compare \texttt{BlockingPy} with \texttt{blocklib} in terms of computational efficiency, accuracy, and number of pairs from the blocking procedures. While \texttt{blocklib} is faster, it results in a significantly higher number of missed pairs and several orders of magnitude more pairs. We also provide a~comparison with the GPU version of the package. 

The only software that provides similar functionality is the \texttt{blocklib} package (cf. \cite{zhang2018scalable, Anonlink}), \texttt{vicinity} \cite{vicinity}, and \texttt{pyJedAI}. The first uses probability signature and LSH-based Lambda-Fold Redundant as described in \cite{zhang2018scalable}. The second uses \texttt{semhash} \cite{minishlab2025semhash}, which allows deduplication of records using kNN-based embedding comparisons and offers exact-match blocking. The third provides an end-to-end entity resolution pipeline that includes multiple key-based blocking methods and comparison cleaning (meta-blocking) strategies, with an optional embedding-based candidate generation component based on \texttt{FAISS}. In contrast, \texttt{BlockingPy} allows various inputs as well as ANN algorithms and graphs, focusing on providing accurate blocking for probabilistic record linkage and deduplication. In \ref{appen-blocklib}, we compare \texttt{BlockingPy} with \texttt{blocklib} in terms of computational efficiency, accuracy, and number of pairs from the blocking procedures. While \texttt{blocklib} is faster, it results in a significantly higher number of missed pairs and several orders of magnitude more pairs. We also provide a~comparison with the GPU version of the package. In \ref{appen-pyjedai} we compare \texttt{PyJedAI} with \texttt{BlockingPy} on the \texttt{Abt-Buy} benchmark \cite{koepcke2010evaluation}.

\section{Software description}\label{sec-software}

\subsection{Installation}

\texttt{BlockingPy} is a Python package released under the MIT license. 
It can be installed from PyPI using package managers such as \texttt{pip} or \texttt{Poetry}.

\begin{lstlisting}[language=Bash]
pip install blockingpy
poetry add blockingpy
\end{lstlisting}

\noindent\textbf{GPU option.} We also provide \texttt{blockingpy-gpu} with CUDA-enabled GPU execution:

\begin{lstlisting}[language=Bash]
# 1) Create env
mamba create -n blockingpy-gpu python=3.10 -y
conda activate blockingpy-gpu

# 2) Install FAISS GPU (tested cuVS nightly)
mamba install -c pytorch/label/nightly \
  faiss-gpu-cuvs=1.11.0=py3.10_ha3bacd1_55_cuda12.4.0_nightly -y

# 3) Install BlockingPy (GPU build)
pip install blockingpy-gpu
\end{lstlisting}

\noindent\textit{Notes:} Import remains \texttt{import blockingpy}.

\subsection{Software functionalities}
% \textit{  Present the major functionalities of the software.}
This subsection describes the main functionalities provided by the \texttt{BlockingPy} package.
\subsubsection{Blocking records}

The core functionality of \texttt{BlockingPy} is to provide an efficient and scalable approach for blocking (also known as indexing) records in both record linkage and deduplication pipelines. Scalability is achieved mainly due to the ANN libraries that we use in our implementations. Users can input either previously computed Document-Term Matrices (DTMs) or raw text data, which the package transforms into DTMs by constructing character $n$-gram representations of the input. Alternatively, users can represent the input data as vector embeddings. For models supported by the \texttt{model2vec} library, users can simply pass the model identifier as a string (e.g., \texttt{"minishlab/potion-base-8m"}), and the package computes the embeddings internally. For any other embedding model, users can precompute the embeddings separately and pass them to the package as NumPy arrays.

Subsequently, similarity-based nearest neighbour search is performed for each record in either the input dataset (deduplication) or query dataset (record linkage). This step is accomplished using one of the following Python implementations of ANN algorithms:

\begin{itemize}
    \item Exact k-Nearest Neighbours (KNN) from \texttt{mlpack} and \texttt{faiss} packages.
    \item Locality Sensitive Hashing (LSH) from \texttt{mlpack} \cite{mlpack2023} and \texttt{faiss} packages.
    \item Hierarchical Navigable Small Worlds (HNSW) from \texttt{faiss} \cite{douze2024faiss}, \texttt{hnswlib} \cite{malkov2018efficient}, and \texttt{voyager} \cite{spotify_voyager} packages.
    \item Nearest Neighbour Descent (NND) from \texttt{pynndescent} \cite{mcinnes_pynndescent} package.
    \item Random projections and NN trees from the \texttt{annoy} \cite{spotify_annoy} package.
    \item Inverted File Flat (IVF), IVF with Product Quantization (IVFPQ) and CAGRA \cite{Ootomo2024CAGRA} from \texttt{faiss} (GPU) package.
\end{itemize}

% Users can fine-tune each algorithm according to their needs. Finally, \texttt{BlockingPy} groups records into blocks by identifying connected components in an undirected graph (using the \texttt{igraph} package \cite{csardi2006igraph}), generated by previous neighbourhood search results. By using the \texttt{block()} method with the desired parameters on the \texttt{Blocker} instance, users can run the above-mentioned process to obtain blocking results.

Users can fine-tune each algorithm according to their needs. Finally, \texttt{BlockingPy} groups records into blocks by identifying connected components in an undirected graph (using the \texttt{igraph} package \cite{csardi2006igraph}), generated by previous neighbourhood search results. In this graph, vertices represent records and edges connect pairs of records identified as neighbours. This step is necessary to handle transitive relationships (e.g., record A matches record B and record B matches record C).  By using the \texttt{block()} method with the desired parameters on the \texttt{Blocker} instance, users can run the above-mentioned process to obtain blocking results.

\subsubsection{Evaluation of blocking result}

The blocking quality can be evaluated by providing ground truth blocks to either the \texttt{block()} or \texttt{eval()} methods. This allows users to assess the quality of the blocks computed by \texttt{BlockingPy}. The key evaluation information generated by the package is described in the sections below.

\vspace{1em} \textit{Reduction Ratio.} Provides necessary details about the reduction in comparison pairs if the given blocks are applied to a further record linkage or deduplication procedure. For deduplication:

$$
\text{RR}_{\text{deduplication}} = 1 - \frac{\sum\limits_{i=1}^{k} \binom{|B_i|}{2}}{\binom{n}{2}},
$$

where $k$ is the total number of blocks, $n$ is the total number of records in the dataset, and $|B_i|$ is the number of records in the $i$-th block. $\sum\limits_{i=1}^{k} \binom{|B_i|}{2}$ is the number of comparisons after blocking, while $\binom{n}{2}$ is the total number of possible comparisons without blocking. For record linkage the reduction ratio is defined as follows

$$
\text{RR}_{\text{record\_linkage}} = 1 - \frac{\sum\limits_{i=1}^{k} |B_{i,x}| \cdot |B_{i,y}|} {(m \cdot n)},
$$

where $m$ and $n$ are the sizes of datasets $X$ and $Y$, and $k$ is the total number of blocks. The term $|B_{i,x}|$ is the number of unique records from dataset $X$ in the $i$-th block, while $|B_{i,y}|$ is the number of unique records from dataset $Y$ in the $i$-th block. The expression $\sum\limits_{i=1}^{k} |B_{i,x}| \cdot |B_{i,y}|$ is the number of comparisons after blocking.

Confusion matrix presents results in comparison to ground-truth \texttt{blocks} in a pairwise manner (e.g., one true positive pair occurs when both records from the comparison pair belong to the same predicted \texttt{block} and to the same ground-truth \texttt{block} in the evaluation data frame).

\begin{itemize}
\item True Positive (TP): Record pairs correctly matched in the same block.
\item False Positive (FP): Records pairs identified as matches that are not true matches in the same block.
\item True Negative (TN): Record pairs correctly identified as non-matches (different blocks)
\item False Negative (FN): Records identified as non-matches that are true matches in the same block.
\end{itemize}

Finally, table \ref{tab-classification-metrics} presents metrics that enables users to evaluate blocking procedures.

\begin{table}[H]
  \centering
  \resizebox{\textwidth}{!}{
    \begin{tabular}{@{}ll|ll@{}}
      \toprule
      \textbf{Metric} & \textbf{Formula} & \textbf{Metric} & \textbf{Formula} \\ 
      \midrule
      Recall (Pairs Completeness) & $\displaystyle \frac{TP}{TP + FN}$ & 
      Accuracy & $\displaystyle \frac{TP + TN}{TP + TN + FP + FN}$ \\[8pt]
      
      Precision (Pairs Quality) & $\displaystyle \frac{TP}{TP + FP}$ & 
      Specificity & $\displaystyle \frac{TN}{TN + FP}$ \\[8pt]
      
      F1 Score & $\displaystyle 2 \cdot \frac{\text{Precision} \times \text{Recall}}{\text{Precision} + \text{Recall}}$ & 
      False Positive Rate & $\displaystyle \frac{FP}{FP + TN}$ \\[8pt]
      
      False Negative Rate & $\displaystyle \frac{FN}{FN + TP}$ & & \\
      \bottomrule
    \end{tabular}
  }
  \caption{Evaluation Metrics}
  \label{tab-classification-metrics}
\end{table}

\subsection{Software architecture}

\begin{figure}[ht!]
    \centering
    \includegraphics[width=0.7\textwidth]{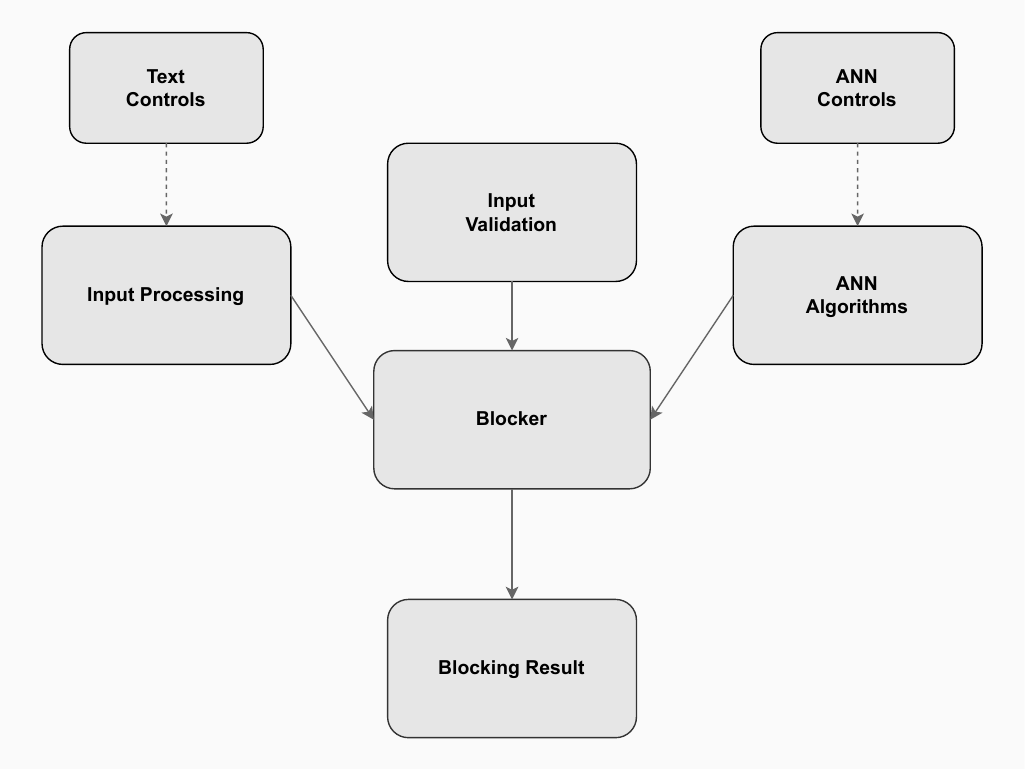}
    \caption{The Architecture of the \texttt{BlockingPy} package}
    \label{fig:architecture}
\end{figure}

The architecture of \texttt{BlockingPy} provides a modular design, where each component is responsible for a specific aspect of the entire blocking workflow. This approach facilitates the integration of new ANN algorithms and modifications to existing functionality. The main components are illustrated in \hyperref[fig:architecture]{Figure~\ref*{fig:architecture}}, and their roles are described below.

\vspace{1em} \textit{Blocker}. The primary component of the \texttt{BlockingPy} architecture. It coordinates interactions between other components, managing the whole workflow of the package from handling data ingestion to generating results.

\vspace{1em} \textit{ANN Algorithms}. This is a collection of ANN implementations with a common interface for the straightforward addition of new algorithms.
The parameters of each algorithm can be adjusted by providing the \texttt{Controls} dictionary to the \texttt{control\_ann} parameter within the \texttt{block} method. An example of this is shown in \hyperref[lst:ann_algo]{Listing~\ref*{lst:ann_algo}}. All ANN backends share one interface across CPU and GPU.

\begin{lstlisting}[label={lst:ann_algo},caption={Setting the algorithm and fine-tuning with control\_ann parameter}]
blocker = Blocker()
result = blocker.block(x=x, ann='voyager', control_ann=control_ann)
\end{lstlisting}

\vspace{1em} \textit{Input Processing}. Manages the data pre-processing and transformation across multiple input formats. Converts dense matrices, sparse matrices, and raw text data to \texttt{Numpy} arrays or \texttt{Scipy} CSR matrices via the \texttt{DataHandler} class for efficient processing. Currently, we support transformation based on $n$-grams and vector embeddings.

\vspace{1em} \textit{Controls}. Manages the configuration system for both text and ANN parameters. Includes customization for algorithm-specific options e.g., metric selection, search and query parameters, alongside tokenization rules and $n$-gram size, or alternatively the choice of an embedding model. \hyperref[lst:controls]{Listing~\ref*{lst:controls}} presents an example of a~\texttt{Controls} dictionary for the \texttt{hnsw} algorithm.

\begin{lstlisting}[label={lst:controls},caption={Controls for \texttt{hnsw} algorithm}]
control_ann = {
    'random_seed': 2025,
    'hnsw': {
        'distance': 'cosine',
        'k_search': 30,
        'n_threads': 1,   
        'path': None,   
        'M': 25,           
        'ef_c': 200,  
        'ef_s': 200,       
    }
}
\end{lstlisting}

\vspace{1em} \textit{Input Validation}. Ensures validity of the data and parameters while providing error messages for the user.

\vspace{1em} \textit{Blocking Result}. The component responsible for managing results and the corresponding information about the blocking process e.g., algorithm used, evaluation metrics, reduction ratio, and block distribution.

\section{Illustrative examples}\label{sec-examples}

In the following section, we demonstrate the \texttt{BlockingPy} package's functionality using two examples: one for record linkage and one for deduplication. We aim to show the ease of use and effectiveness of the package in blocking records for entity resolution.

\subsection{Record linkage}

In this example, we use the Cis-Census datasets (for more information see \cite{McLeod2011CensusCIS}) to show the record linkage capabilities of \texttt{BlockingPy}.
These datasets contain fictitious personal information, with the \texttt{census} dataset comprising 25,343 records and the \texttt{cis} (Customer Information System) dataset containing 24,613 records.
\hyperref[tab:census_data]{Table~\ref{tab:census_data}} shows sample records from the \texttt{census} dataset after the initial preparation described in \hyperref[lst:rl_pre]{Listing~\ref*{lst:rl_pre}}. The \texttt{cis} dataset is structured similarly. 

\begin{table}[ht!]
    \centering
    \scriptsize
    \resizebox{\textwidth}{!}{ 
    \begin{tabular}{l l c c c c l l} 
        \toprule
        \textbf{NAME1} & \textbf{NAME2} & \textbf{SEX} & \textbf{D} & \textbf{M} & \textbf{Y} & \textbf{ENUMCAP} & \textbf{ENUMPC} \\
        \midrule
        COUIE  & PRICE & M & 1  & 6  & 1960 & 1 WINDSOR ROAD & DE03US \\
        ABBIE  & PVICE & F & 9  & 11 & 1961 & 1 WINDSOR ROAD & DE03US \\
        LACEY  & PRICE & F & 7  & 2  & 1999 & 1 WINDSOR ROAD & DE03US \\
        SAMUEL & PRICE & M & 13 & 4  & 1990 & 1 WINDSOR ROAD & DE03US \\
        JOSEPH & PRICE & M & 20 & 4  & 1986 & 1 WINDSOR ROAD & DE03US \\
        \bottomrule
    \end{tabular}
    }
        \caption{Example records from \texttt{census} dataset with variables used for \texttt{txt} column. Variables (changed here for better printing): \textbf{NAME1} (first name), \textbf{NAME2} (surname), \textbf{SEX} (gender), \textbf{D}, \textbf{M}, \textbf{Y} (birth date), \textbf{ENUMCAP} (address), \textbf{ENUMPC} (postal code).}

    \label{tab:census_data}
\end{table}

\begin{lstlisting}[caption=Preprocessing the record linkage example data, label=lst:rl_pre]
from blockingpy import Blocker
import pandas as pd
from blockingpy.datasets import load_census_cis_data

census, cis = load_census_cis_data()

census = census[["PERSON_ID","PERNAME1","PERNAME2","SEX","DOB_DAY","DOB_MON","DOB_YEAR","ENUMCAP","ENUMPC"]]
cis = cis[["PERSON_ID","PERNAME1","PERNAME2","SEX", "DOB_DAY","DOB_MON","DOB_YEAR","ENUMCAP","ENUMPC"]]
\end{lstlisting}

Before blocking, we merge all fields of interest (without \texttt{ID}) into a single field (i.e., \texttt{txt}) that we will pass to the algorithm as shown in \hyperref[lst:rl_txt]{Listing~\ref*{lst:rl_txt}} (i.e, we use \textit{schema-agnostic} blocking). This step is performed for both datasets.

\begin{lstlisting}[caption=Creating the \texttt{txt} column, label=lst:rl_txt]
for df in [census, cis]:
    df[['DOB_DAY', 'DOB_MON', 'DOB_YEAR']] = (
        df[['DOB_DAY', 'DOB_MON', 'DOB_YEAR']]
        .astype("Int64")
        .astype(str)
        .replace('<NA>', '')
    )
    df.fillna('', inplace=True)
    
    df['txt'] = (
        df['PERNAME1']
        + df['PERNAME2']
        + df['SEX']
        + df['DOB_DAY']
        + df['DOB_MON']
        + df['DOB_YEAR']
        + df['ENUMCAP']
        + df['ENUMPC']
    )
\end{lstlisting}

Subsequently, in \hyperref[lst:rl_block]{Listing~\ref*{lst:rl_block}} we initialise \texttt{Blocker}, invoke the \texttt{block()} method, and select the \texttt{hnsw} algorithm with default parameters to obtain the results. The tokenization and white-space removal are handled by the algorithm internally and can be modified via the \texttt{control\_txt} parameter. Additional algorithm-specific parameters can be fine-tuned via the \texttt{control\_ann} argument.

\begin{lstlisting}[caption=Record linkage blocking example, label=lst:rl_block]
blocker = Blocker()
rec_lin_result = blocker.block(
    x=census['txt'],
    y=cis['txt'],
    ann='hnsw',
    random_seed=42
)
\end{lstlisting}

The output is a data frame containing rows with indices from both datasets (\texttt{x}, \texttt{y}) alongside the assigned \texttt{block} (an integer starting with 0) and distance computed with one of the available metrics. \hyperref[lst:rlResult]{Listing~\ref{lst:rlResult}} presents a snippet of the output in this example. \hyperref[lst:rlBlockInfo]{Listing~\ref*{lst:rlBlockInfo}} shows the basic blocking information logged by the package. The first section provides details about the algorithm used, the number of generated blocks, the number of columns used for blocking (created by shingling, a preprocessing step that converts strings into overlapping letter groups used as input to the ANN algorithm), and the reduction ratio. The second section displays the distribution of block sizes; for example, there were $591$ blocks created, each containing $3$ records.

\begin{lstlisting}[caption=Record linkage example of blocking result, label=lst:rlResult]
print(rec_lin_result.result.head())
       x  y  block      dist
0  17339  0      0  0.134151
1   9567  1      1  0.064307
2  10389  2      2  0.044183
3  24258  3      3  0.182125
4   3714  4      4  0.288487
\end{lstlisting}

\begin{lstlisting}[caption=Basic blocking information, label=lst:rlBlockInfo]
print(rec_lin_result)
====================================================
Blocking based on the hnsw method.
Number of blocks: 23993
Number of columns created for blocking: 1072
Reduction ratio: 0.999961
====================================================
Distribution of the size of the blocks:
Block Size | Number of Blocks
         2 | 23388          
         3 | 591            
         4 | 13             
         5 | 1  
\end{lstlisting}

We observe a reduction ratio of $0.9999$, which indicates that the possible number of candidate pairs is substantially reduced, thus lowering the computational complexity, which is crucial for large-scale datasets.

With a unique personal identifier available in both datasets, we can evaluate the algorithm. For that, we create a \texttt{true\_blocks} data frame which represents indices from both datasets and their true block, as shown in \hyperref[lst:rl_tb_creation]{Listing~\ref*{lst:rl_tb_creation}}. In this case, every record in each block should point to the same unique entity. Furthermore, in \hyperref[lst:rl_tb]{Listing~\ref*{lst:rl_tb}} we pass the \texttt{rec\_lin\_result} object (from \hyperref[lst:rl_block]{Listing~\ref*{lst:rl_block}}) and the \texttt{true\_blocks} to the \texttt{eval} method to obtain the evaluation results. In this example, we randomly sampled 1000 ground-truth pairs to showcase the evaluation functionality of the \texttt{BlockingPy} package.

\begin{lstlisting}[caption=Preparation of \texttt{true\_blocks}, label=lst:rl_tb_creation]
census['x'] = range(len(census))
cis['y'] = range(len(cis))

true_blocks = pd.merge(
    left=census[['PERSON_ID', 'x']],
    right=cis[['PERSON_ID', 'y']],
    on='PERSON_ID'
)

true_blocks['block'] = range(len(true_blocks))
\end{lstlisting}

\begin{lstlisting}[caption=Record Linkage evaluation code, label=lst:rl_tb]
true_blocks = true_blocks.sample(1000, random_state=42)
eval_result = blocker.eval(rec_lin_result, true_blocks[['x', 'y', 'block']])
\end{lstlisting}

Through this step, the user can access both the evaluation metrics and the confusion matrix. Evaluation is performed only for records that are available in \texttt{true\_blocks}. We assume that we have no knowledge about the records not included in the ground-truth data frame. \hyperref[lst:rl_metrics]{Listing~\ref*{lst:rl_metrics}} and \hyperref[lst:rl_conf]{Listing~\ref*{lst:rl_conf}} present the metrics and confusion matrix computed in this example.

\begin{lstlisting}[caption=Evaluation Metrics, label=lst:rl_metrics]
print(eval_result.metrics)
recall         0.997000
precision      1.000000
fpr            0.000000
fnr            0.003000
accuracy       0.999997
specificity    1.000000
f1_score       0.998498
\end{lstlisting}

\begin{lstlisting}[caption=Confusion Matrix, label=lst:rl_conf]
print(eval_result.confusion)
               Predicted Positive  Predicted Negative
Actual Positive        997                  3
Actual Negative         0                 999000
\end{lstlisting}

The results present \texttt{BlockingPy's} ability to efficiently and accurately block records for record linkage tasks. The full version of this example is available on the documentation website.

\subsection{Deduplication}

In this example, our aim is to show how \texttt{BlockingPy} can be integrated into full entity resolution workflows. For that, we will present the deduplication of the \texttt{febrl1} dataset obtained from the \texttt{recordlinkage} package \cite{recordlinkage}. This dataset was generated using the \texttt{FEBRL} software and contains 1000 records with fictitious personal information, of which 500 are original and 500 are duplicates with introduced errors. Firstly, we prepare the blocking key by imputing missing values with an empty string and concatenating fields with personal information into a single field, which we will use for blocking. After initialising \texttt{Blocker} and performing blocking on said data, we are given the blocks by \texttt{BlockingPy}. \hyperref[lst:dd_pre]{Listing~\ref*{lst:dd_pre}} presents this process, and the \hyperref[lst:dd_block]{Listing~\ref*{lst:dd_block}} shows the blocking information output and the first five rows of the actual result.

\begin{lstlisting}[caption=Preprocessing and blocking, label=lst:dd_pre]
import recordlinkage
from recordlinkage.datasets import load_febrl1
from blockingpy import Blocker
import pandas as pd
import numpy as np

df = load_febrl1()

df = df.fillna('')
df['txt'] = df['given_name'] + df['surname'] + \
            df['street_number'] + df['address_1'] + \
            df['address_2'] + df['suburb'] + \
            df['postcode'] + df['state'] + \
            df['date_of_birth'] + df['soc_sec_id']

blocker = Blocker()
blocking_result = blocker.block(
    x=df['txt'],
    ann='hnsw',
    random_seed=42
) 
\end{lstlisting}

\begin{lstlisting}[caption=Blocking Information, label=lst:dd_block]
print(blocking_result)
========================================================
Blocking based on the hnsw method.
Number of blocks: 500
Number of columns used for blocking: 1023
Reduction ratio: 0.998999
========================================================
Distribution of the size of the blocks:
Block Size | Number of Blocks
         2 | 500

print(blocking_result.result.head())
    x  y  block      dist
  474  0      0  0.048375
  330  1      1  0.038961
  351  2      2  0.086690
  290  3      3  0.024617
  333  4      4  0.105662
\end{lstlisting}

Both columns \texttt{x} and \texttt{y} refer to indices from the \texttt{febrl1} dataset alongside their designated \texttt{block} and distance measured between them using one of the available metrics (cosine in this example, the default for \texttt{hnsw}). 
Furthermore, the necessary step to make integration with \texttt{recordlinkage} possible is to add the \texttt{block} column to the original data frame, which we can do with the \texttt{add\_block\_column} method as presented in \hyperref[lst:dd_col]{Listing~\ref*{lst:dd_col}}. 

\begin{lstlisting}[caption=Creating the \texttt{block} column, label=lst:dd_col]
df['id'] = range(len(df))
df_final = blocking_result.add_block_column(df, id_col_left='id')
\end{lstlisting}

After calling the \texttt{block} method on the \texttt{Index} with the previously obtained \texttt{block} column in \hyperref[lst:dd_linker]{Listing~\ref*{lst:dd_linker}} we can proceed with the usual \texttt{recordlinkage} workflow. 

The full version of this example and many others, alongside the code, can be found in the documentation.

\begin{lstlisting}[caption=Integration with \texttt{recordlinkage}, label=lst:dd_linker]
indexer = recordlinkage.Index()
indexer.block('block')
pairs = indexer.index(df_final)
\end{lstlisting}

\section{Impact}\label{sec-impact}

The proposed approach is used at Statistics Poland in the procedure for the linkage of records between administrative datasets as well as within one dataset. For instance, we have used this approach to deduplicate records of forced migrants crossing the Polish-Ukrainian border after the full-scale invasion of Ukraine by Russia on February 24, 2022. We also used this technique to determine the number of residents of Ukraine under temporary protection in Poland as of March 31, 2023 \cite{StatisticsPoland2023}. The main motivation behind this was the problem of the lack of identifiers as well as the transliteration from Ukrainian/Russian to Polish language (e.g. \foreignlanguage{ukrainian}{Олександр}, should be Ołeksandr not Aleksandr).  Before implementing this solution, Statistics Poland relied on user-defined rules for linking data without identifiers, which often resulted in imprecise population flow estimates due to missed matches or false positives.

%This software opens up new research questions regarding optimal blocking strategies in administrative data integration, particularly in contexts where language differences and transliteration complexities exist. Researchers can now investigate how different blocking approaches affect linkage quality in multilingual settings, a question that was previously difficult to study systematically without appropriate tools. The software has meaningfully improved the pursuit of quality of official statistics by allowing for more accurate population estimates through better deduplication of records. 

We believe that the proposed software will be of interest to researchers and practitioners who link data without identifiers or would like to apply large language models to detect duplicates but need to reduce costs by providing a small number of comparison buckets.

\section{Conclusions}

The current project is designed to be extensible so that it can accommodate new ANN algorithms and new input data formats. Any contributions are warmly welcome; see  \url{https://github.com/ncn-foreigners/BlockingPy/issues} for a feature request and bug tracker. 

Future updates will focus on adding functionality for privacy-preserving record linkage without directly sharing personally identifiable information.

\section*{Acknowledgements}

We thank the reviewers for their comments on this manuscript. This software was a~part of Tymoteusz Strojny's Bachelor's thesis at Poznań University of Economics and Business, Poland. \texttt{BlockingPy} is based on the  \texttt{blocking} package in R (see \url{https://ncn-foreigners.github.io/blocking/}).  

Work on this package is supported by the National Science Centre, OPUS 20 grant no. 2020/39/B/HS4/00941 (\textit{Towards census-like statistics for foreign-born populations -- quality, data integration and estimation}).

This software was peer-reviewed and accepted by the \texttt{pyOpenSci} organisation (see \url{https://github.com/pyOpenSci/software-submission/issues/232}).

\section*{CRediT authorship contribution statement}

\textit{Tymoteusz Strojny}: Conceptualization, Software, Writing -- Original Draft, Writing -- Review \& Editing;  \textit{Maciej Beręsewicz}: Conceptualization, Writing -- Original Draft, Writing -- Review \& Editing, Supervision, Methodology, Project administration, Validation.

\section*{Declaration of competing interest}

The authors declare that they have no known competing financial interests or personal relationships that could have appeared to influence the work reported in this paper.

\section*{Declaration of generative AI and AI-assisted technologies in the writing process}

During the preparation of this work the author(s) used \texttt{DeepL} and \texttt{Claude.ai} to proofread the paper. After using this  the author(s) reviewed and edited the content as needed and take(s) full responsibility for the content of the published article.

\section*{Data availability}

% Data are publicly available.
The datasets used in Section~\ref{sec-examples} are included with the \texttt{BlockingPy} package and are also publicly available from their original sources.


\clearpage

\appendix

\section{Comparison of blocking methods implemented in \texttt{splink}, \texttt{blocklib}, \texttt{recordLinkage}, \texttt{pyJedi} and \texttt{BlockingPy}}

%% tabela 2 kolumnuy: 1 pakiet, 2 lista metod

\begin{table}[H]
\centering
\caption{Comparison of blocking methods implemented in selected software.}
\label{tab:blocking_methods}

% Increase row height for better readability
\renewcommand{\arraystretch}{1.4} 

\begin{tabularx}{\textwidth}{ @{} l >{\RaggedRight\arraybackslash}X @{} }
\toprule
\textbf{Package} & \textbf{Blocking methods} \\
\midrule

\texttt{splink} & 
Rule-based / deterministic blocking (SQL-style rules), phonetic blocking (e.g., Soundex, Metaphone). \\
\addlinespace

\texttt{blocklib} & 
Probabilistic signature blocking (p-sig), LSH-based \(\lambda\)-fold redundant blocking (Bloom filter + LSH). \\
\addlinespace

\texttt{recordlinkage} & 
Exact blocking, Sorted Neighbourhood, full indexing (all pairs), random indexing (sampled pairs). \\
\addlinespace

\texttt{pyJedAI} & 
Token Blocking, Sorted Neighborhood, Extended Sorted Neighborhood, Q-Grams Blocking, Extended Q-Grams Blocking, Suffix Arrays Blocking, Extended Suffix Arrays Blocking, Embeddings-based NN blocking. \\
\addlinespace

\texttt{BlockingPy} & 
ANN-based blocking via FAISS, HNSW (faiss / hnswlib / voyager), NN-Descent (pynndescent), Annoy, plus GPU indices (e.g., IVF, IVFPQ, CAGRA); blocks from connected components of the NN graph. \\

\bottomrule
\end{tabularx}
\end{table}

\clearpage
\section{Numerical comparison with \texttt{blocklib}}\label{appen-blocklib}

We compare \texttt{BlockingPy} and \texttt{blocklib} in three synthetic datasets ($1,500$, $15,000$, and $150,000$ records) with duplicates $500$, $5,000$, and $50,000$, respectively. The results are averaged over 10 runs. The datasets were generated with \texttt{geco3} and contain fictional personal information (given name, middle name, surname, municipality, date of birth, and nationality). CPU experiments used a 6-core Intel Core i5 with 16\,GB RAM and Python~3.12. GPU runs used a NVIDIA RTX~3050 (4\,GB VRAM) and Python~3.10. We report time (s), recall (Pairs Completeness), reduction ratio (RR), and candidate pairs (millions). The comparison is also available in the documentation along with the code and data in the project repository. Preprocessing: for \texttt{BlockingPy}, we concatenate all fields of interest into a single text field; for \texttt{blocklib}, we tune algorithm parameters to approximately match the reduction ratio of the other methods to ease comparison.

\begin{table}[H]
  \centering
  \caption{CPU results on three datasets.}
  \label{tab:cpu-blocklib-vs-blockingpy}
  \setlength{\tabcolsep}{6pt}
  \resizebox{\textwidth}{!}{%
  \begin{tabular}{lrrrrr}
    \toprule
    Algorithm & \multicolumn{1}{c}{Size} & \multicolumn{1}{c}{Time [s]} & \multicolumn{1}{c}{Recall} & \multicolumn{1}{c}{RR} & \multicolumn{1}{c}{Pairs (M)} \\
    \midrule
P-Sig & 1,500 & 0.051 (0.010) & 0.599 (0.000) & 0.996 (0.000) & 0.004 (0.000) \\
$\lambda$-fold LSH & 1,500 & 0.193 (0.009) & 0.466 (0.033) & 0.991 (0.003) & 0.010 (0.003) \\
BlockingPy (voyager) & 1,500 & 0.293 (0.029) & 0.952 (0.006) & 0.997 (0.000) & 0.003 (0.000) \\
BlockingPy (faiss\_hnsw) & 1,500 & 0.164 (0.049) & 0.960 (0.000) & 0.998 (0.000) & 0.003 (0.000) \\
BlockingPy (faiss\_lsh) & 1,500 & 0.185 (0.015) & 0.955 (0.008) & 0.997 (0.000) & 0.003 (0.000) \\
    \midrule
P-Sig & 15,000 & 0.388 (0.087) & 0.616 (0.000) & 0.996 (0.000) & 0.407 (0.000) \\
$\lambda$-fold LSH & 15,000 & 2.016 (0.251) & 0.454 (0.028) & 0.992 (0.003) & 0.940 (0.337) \\
BlockingPy (voyager) & 15,000 & 6.777 (0.903) & 0.876 (0.005) & 1.000 (0.000) & 0.041 (0.001) \\
BlockingPy (faiss\_hnsw) & 15,000 & 12.029 (1.681) & 0.913 (0.000) & 1.000 (0.000) & 0.031 (0.000) \\
BlockingPy (faiss\_lsh) & 15,000 & 1.398 (0.148) & 0.900 (0.004) & 1.000 (0.000) & 0.033 (0.001) \\
    \midrule
P-Sig & 150,000 & 3.444 (0.091) & 0.609 (0.000) & 0.996 (0.000) & 40.231 (0.000) \\
$\lambda$-fold LSH & 150,000 & 19.848 (0.476) & 0.450 (0.026) & 0.992 (0.003) & 95.064 (34.252) \\
BlockingPy (voyager) & 150,000 & 110.341 (3.815) & 0.715 (0.005) & 1.000 (0.000) & 0.675 (0.021) \\
BlockingPy (faiss\_hnsw) & 150,000 & 250.012 (4.232) & 0.832 (0.001) & 1.000 (0.000) & 0.375 (0.001) \\
BlockingPy (faiss\_lsh) & 150,000 & 56.544 (1.315) & 0.818 (0.001) & 1.000 (0.000) & 0.400 (0.005) \\
    \bottomrule
  \end{tabular}
  }
\end{table}

\begin{table}[H]
  \centering
  \caption{GPU results on 150,000 records dataset.}
  \label{tab:gpu}
  \setlength{\tabcolsep}{6pt}
    \resizebox{\textwidth}{!}{%
  \begin{tabular}{lrrrrr}
    \toprule
    Algorithm & \multicolumn{1}{c}{Size} & \multicolumn{1}{c}{Time [s]} & \multicolumn{1}{c}{Recall} & \multicolumn{1}{c}{RR} & \multicolumn{1}{c}{Pairs (M)} \\
    \midrule
BlockingPy (gpu\_faiss cagra) & 150,000 & 51.084 (0.802) & 0.827 (0.000) & 1.000 (0.000) & 0.380 (0.000) \\
BlockingPy (gpu\_faiss flat) & 150,000 & 22.511 (0.508) & 0.839 (0.000) & 1.000 (0.000) & 0.367 (0.000) \\
BlockingPy (gpu\_faiss ivf) & 150,000 & 71.458 (2.185) & 0.801 (0.004) & 1.000 (0.000) & 0.487 (0.021) \\
    \bottomrule
  \end{tabular}
  }
\end{table}


%% References:

\clearpage
\section{Comparison with \texttt{PyJedAI}}\label{appen-pyjedai}

We compare \texttt{BlockingPy} and \texttt{PyJedAI} on the Abt-Buy clean-clean entity resolution benchmark, which consists of two product catalogs from the online retailers Abt.com and Buy.com. We use the Abt-Buy dataset as provided in the \texttt{pyJedAI} repository, which contains 1{,}076 records in each source and a gold-standard mapping between the two sources. The hardware setup is the same as in \ref{appen-blocklib}; the experiments were run with Python~3.11 (with \texttt{PyJedAI} using NumPy~2.4.1 and \texttt{BlockingPy} using NumPy~1.26.4). We report runtime (s), recall, precision, $F_1$, and the number of candidate comparison pairs.

For \texttt{BlockingPy}, we use the HNSW backend with the \texttt{minishlab/potion-base-32M} embedding model; all remaining parameters are left at their defaults. For \texttt{PyJedAI}, we use the \texttt{sminilm} vectorizer with $k=5$ nearest neighbors and Euclidean distance. The data (and an example workflow) are available in the \texttt{pyJedAI} documentation/repository. The \texttt{BlockingPy} workflow used for this comparison is likewise available in the \texttt{BlockingPy} repository.

\begin{table}[H]
  \centering
  \caption{Comparison on the \texttt{Abt-Buy} benchmark}
  \label{tab:pyjedai}
  \setlength{\tabcolsep}{6pt}
  \begin{adjustbox}{max width=\textwidth}
    \begin{tabular}{lrrrrr}
      \toprule
      Algorithm & \multicolumn{1}{c}{Time [s]} & \multicolumn{1}{c}{Recall} & \multicolumn{1}{c}{Precision} & \multicolumn{1}{c}{$F_1$} & \multicolumn{1}{c}{Pairs}\\
      \midrule
      PyJedAI    & 52.6 & 0.9377 & 0.1875 & 0.3126 & 5380 \\
      BlockingPy & 4.4  & 0.8234 & 0.8234 & 0.8234 & 1076 \\
      \bottomrule
    \end{tabular}
  \end{adjustbox}
\begin{flushleft}
\small
\justifying
\textit{Note}: The identical recall, precision, and $F_1$ values reported here are expected in this setting and are not a reporting error: in \texttt{BlockingPy} we use $k=1$ (one candidate per record) and, for this experiment, the graph-based step did not find additional links, so evaluation is effectively performed on the same one-to-one pair set; under these conditions these three metrics coincide.
\end{flushleft}
\end{table}

\clearpage

\bibliographystyle{elsarticle-num} 
\bibliography{bibliography}


% \textit{If the software repository you used supplied a DOI or another
% Persistent IDentifier (PID), please add a reference for your software
% here. For more guidance on software citation, please see our guide for
% authors or \href{https://f1000research.com/articles/9-1257/v2}{this
%   article on the essentials of software citation by FORCE 11}, of
% which Elsevier is a member.}

% \large{\textbf{Reminder: Before you submit, please delete all 
% the instructions in this document, 
% including this paragraph. 
% Thank you!}}

\end{document}